\documentclass[fleqn,10pt]{wlscirep}
\usepackage[utf8]{inputenc}
\usepackage[T1]{fontenc}

\title{GazeBase: A Large-Scale, Multi-Stimulus, Longitudinal Eye Movement Dataset}

\author[1,*]{Henry Griffith}
\author[1]{Dillon Lohr}
\author[1]{Evgeny Abdulin}
\author[1]{Oleg Komogortsev}
\affil{Texas State University, Department of Computer Science, San Marcos, TX, 78666, USA}

\affil[*]{corresponding author: Henry Griffith (h\_169@txstate.edu)}

\newcommand{\figurerefname}{Fig.}  
\newcommand{\figuresrefname}{Figs.}  

\usepackage[figurename=\figurerefname,labelsep=colon]{caption}

\begin{abstract}
This manuscript presents GazeBase, a large-scale longitudinal dataset containing 12,334 monocular eye-movement recordings captured from 322 college-aged subjects. Subjects completed a battery of seven tasks in two contiguous sessions during each round of recording, including a - 1) fixation task, 2) horizontal saccade task, 3) random oblique saccade task, 4) reading task, 5/6) free viewing of cinematic video task, and 7) gaze-driven gaming task. A total of nine rounds of recording were conducted over a 37 month period, with subjects in each subsequent round recruited exclusively from the prior round. All data was collected using an EyeLink 1000 eye tracker at a 1,000 Hz sampling rate, with a calibration and validation protocol performed before each task to ensure data quality. Due to its large number of subjects and longitudinal nature, GazeBase is well suited for exploring research hypotheses in eye movement biometrics, along with other emerging applications applying machine learning techniques to eye movement signal analysis. 

\end{abstract}
\begin{document}

\flushbottom
\maketitle

\thispagestyle{empty}


\section*{Background \& Summary}

Due to their demonstrated uniqueness and persistence \cite{bargary2017individual}, human eye movements are a desirable modality for biometric applications \cite{jain2015guidelines}. Since their original consideration in the early 2000s \cite{kasprowski2004eye}, eye movement biometrics have received substantial attention within the security literature \cite{katsini2020role}. Recent interest in this domain is accelerating, due to the proliferation of gaze tracking sensors throughout modern consumer products, including automotive interfaces, traditional computing platforms, and head-mounted devices for virtual and augmented reality applications. Beyond this increase in sensor ubiquity, eye movements are an advantageous  modality for emerging biometric systems due to their ability to support continuous authentication \cite{eberz2016looks} and liveliness detection \cite{komogortsev2015attack}, along with their ease of fusion with other appearance-based traits in both the eye \cite{winston2019comprehensive} and periocular region \cite{woodard2010periocular}.

Despite considerable research progress in eye movement biometrics over the past two decades, several open research areas remain. Namely, ensuring performance robustness with respect to data quality, and further investigating both task dependency and requisite recording duration is necessary to transition this technology to widespread commercial adoption. Moreover, the exploration of emerging deep learning techniques, which have proven successful for more traditional biometric modalities \cite{sundararajan2018deep}, has been limited for eye movement biometrics. This investigation is impeded by the challenges associated with the large-scale collection of eye movement data, along with the lack of  task-diverse, publicly-available, large-scale data repositories. 

\begin{figure}[ht!]  
\centering
\includegraphics[width=\linewidth]{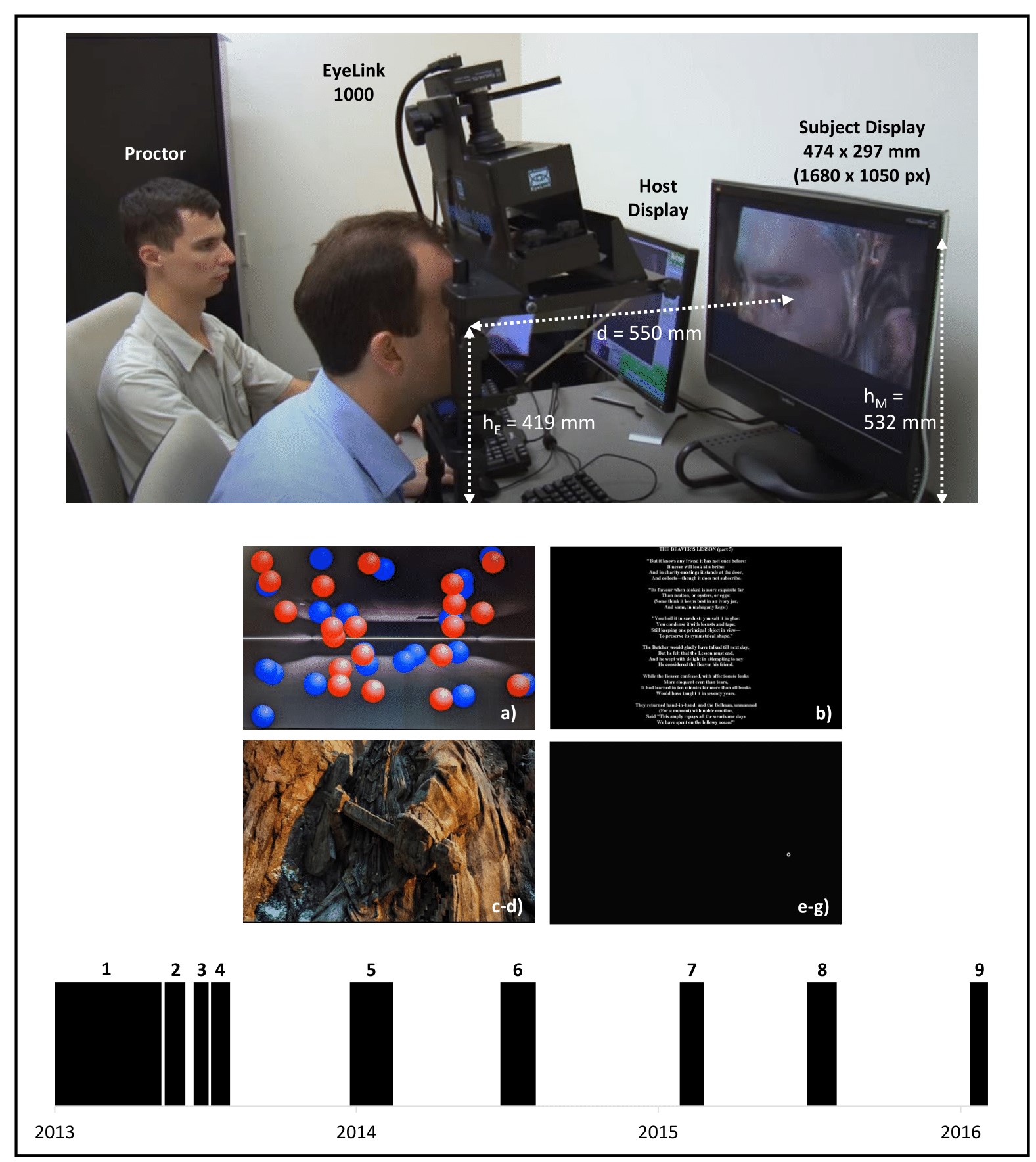}  
\caption{Summary of the GazeBase dataset collection.  \textbf{ \textit{Top:}} Experimental set-up. \textbf{ \textit{Middle:}} Screenshots of the stimuli from four of the seven tasks. a) is a screenshot of the gaze-driven gaming task, b) is a screenshot of the reading task, c-d) is a single screenshot from one of the two video viewing tasks. e-g) shows the stationary bull's-eye target utilized in the calibration and validation process, along with the fixation and two saccade tasks. The screenshot is obtained during the random saccade task. \textbf{ \textit{Bottom:}} A timeline of the multiple recording rounds (round identifiers are labeled on top of each rectangle).}

\label{fig:SummaryFig}
\end{figure}

To promote further development in eye movement biometrics research, this manuscript describes a newly-released dataset consisting of 12,334 monocular eye movement recordings captured from 322 individuals while performing seven discrete tasks. The considered task battery includes guided stimuli intended to induce specific eye movements of interest, along with multiple objective-oriented and free-viewing tasks, such as reading, movie viewing, and game playing. Hereby denoted as GazeBase, this data were captured over a 37 month period during nine rounds of recording, with two contiguous sessions completed during each recording period. The data collection workflow is summarized in \figurerefname~\ref{fig:SummaryFig}.


Although subsets of this data have been utilized in prior work\cite{abdulin2017method, friedman2018novel, rigas2018study, friedman2019assessment, lohr2019evaluating, friedman2020temporal, friedman2017method, griffith2018towards,griffith2020prediction,griffith2018towards1,griffith2020shift,griffith2020texture}, this recent dissemination is the first release of the entire set of gaze recordings and corresponding target locations for applicable stimuli. As the experimental parameters of the data collection were chosen to maximize the utility of the resulting data for biometric applications, GazeBase is well suited for supporting further investigation of emerging machine learning biometric techniques to the eye movement domain, such as metric learning \cite{abdelwahab2019deep}. Beyond this target application, the resulting dataset is also useful for exploring numerous additional research hypotheses in various areas of interest, including eye movement classification and prediction. Applications employing machine learning techniques will benefit both from the scale of available data, along with the diversity in tasks considered and subjects recorded. Moreover, this dissemination will help improve quality in subsequent research by providing a diverse set of recordings for benchmarking across the community \cite{jain2015guidelines}. 

\section*{Methods}
\subsection*{Subjects}
\begin{figure}[ht]
\centering
\includegraphics[width=0.7\linewidth]{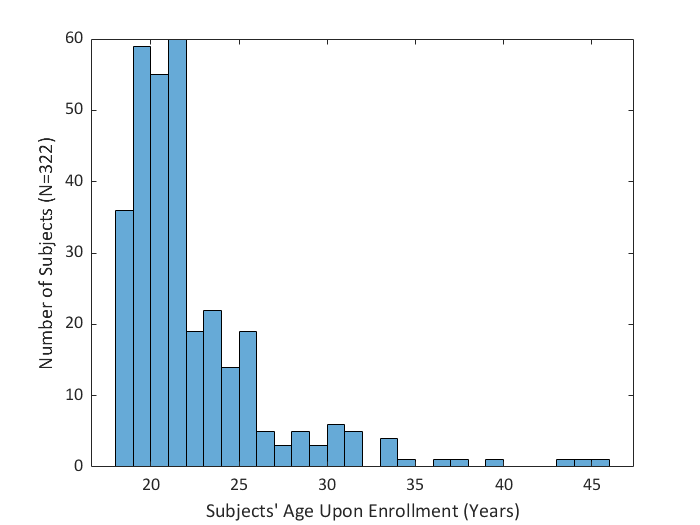}
\caption{Distribution of subjects' ages at the time of enrollment}
\label{fig:AgeDist}
\end{figure}

Subjects were initially recruited from the undergraduate student population at Texas State University through email and targeted in-class announcements. A total of 322 subjects (151 self-identifying as female, 171 self-identifying as male) were enrolled in the study and completed the Round 1 collection in its entirety. Subjects for Rounds 2 - 9 were recruited exclusively from the prior round's subject pool. All subjects had normal or corrected to normal visual acuity. Aggregate subject demographic information is presented in Table~\ref{tab:DemographicInfo}. The distribution of subjects' age at the time of the Round 1 collection is shown in \figurerefname~\ref{fig:AgeDist}. The number of subjects completing the entire task battery in each round is summarized in Table~\ref{tab:RoundInfo}, along with the recording dates for each round of collection. 

\begin{table}[ht]
\centering
\begin{tabular}{llllll}
\toprule
Ethnicity: & Asian & Black & Caucasian & Hispanic & Mixed \\
\midrule
Num. of Subjects: &10 & 32 & 178 & 76 & 27 \\
\bottomrule
\end{tabular}
\caption{Self-reported ethnicity of subjects}
\label{tab:DemographicInfo}
\end{table}

\begin{table}[ht]
\centering
\begin{tabular}{lll}
\toprule
Round ID & Num. of Subjects & Date Range \\
\midrule 
1 & 322 & 09/13 - 02/14\\ 
2 & 136 & 02/14 - 03/14\\ 
3 & 105 & 03/14 - 04/14\\ 
4 & 101 & 04/14 - 04/14\\ 
5 & 78 & 09/14 - 11/14\\ 
6 & 59 & 03/15 - 05/15\\
7 & 35  & 10/15 - 11/15\\ 
8 & 31  & 03/16 - 05/16\\ 
9 & 14  & 10/16 - 11/16\\
\bottomrule
\end{tabular}
\caption{Total number of subjects and recording date range of each round}
\label{tab:RoundInfo}
\end{table}

 All subjects provided informed consent under a protocol approved by the Institutional Research Board at Texas State University prior to each round of recording. As part of the consent process, subjects acknowledged that the resulting data may be disseminated in a de-identified form.  

\subsection*{Data Acquisition Overview}
Data was captured under the supervision of a trained experimental proctor. Before initiating the recording process, the proctor provided a general overview of the experiment to the subject, along with a summary of best practices for maximizing the quality of the captured data. Namely, subjects were instructed to maintain a stable head and body position, and to attempt to avoid excessive blinking. Based upon initial recording experiences, it was ultimately suggested that subjects avoid wearing mascara to the recording session as part of the appointment confirmation email. This suggestion was initiated during the first round of recording and maintained throughout the remainder of the collection. 

Subjects wearing eyeglasses were asked to attempt the experiment with glasses removed. This protocol was chosen due to the known challenges associated with recording individuals wearing eyeglasses using the target capture modality. If subjects were unable to complete the experimental protocol with glasses removed, an attempt was made to complete the experiment while wearing eyeglasses. Subjects that could not be successfully calibrated or recorded after multiple attempts were withdrawn from the study. Subjects could also self-withdrawal at any point during the recording process. A total of 13 subjects were withdrawn from the initial round of the study. GazeBase contains only data from subjects completing the entire recording protocol for a given round. 

Monocular (left) eye movements were captured at a 1,000 Hz sampling rate using an EyeLink 1000 eye tracker (SR Research, Ottawa, Ontario, Canada) in a desktop mount configuration. The EyeLink 1000 is a video oculography device which operates using the pupil-corneal reflection principle, where gaze locations are estimated from pupil-corneal reflection vectors using a polynomial mapping developed during calibration \cite{sr2010eyelink}. Stimuli were presented to the subject on a 1680 x 1050 pixel (474 x 297 mm) ViewSonic (ViewSonic Corporation, Brea, California, USA) monitor. Instrumentation control and recording monitoring were performed by the proctor using a dedicated host computer as shown in \figurerefname~\ref{fig:SummaryFig}. Recordings were performed in a quiet laboratory environment without windows. Consistent ambient lighting was provided by ceiling-mounted fluorescent light fixtures. 

Subjects were seated 550 mm in front of the display monitor. Subjects' heads were stabilized using a chin and forehead rest. Once the participant was initially seated, the chin rest was adjusted to level the subjects' left eye at the primary gaze position, located 36 mm above the center of the monitor. This vertical offset from the monitor center was chosen to ensure the comfort of taller participants given restrictions on adjusting the chair and monitor height. Chair height was initially adjusted as necessary to ensure the comfort of the subject, followed by additional fine tuning of the chin rest as required to align the left eye with the primary gaze position. The lens focus was manually tuned as necessary in order to ensure the sharpness of the eye image as viewed by the proctor on the host display. 

Subjects completed two sessions of recording for each round of collection. While the proctor suggested that subjects take a five-minute break between sessions if needed, subjects were free to decline the break if desired. Subjects could also request breaks at any time during the recording process as noted during the consent process. During each recording, the gaze location was monitored by the proctor to ensure compliance with the individual task protocols. 

The gaze position and corresponding stimuli were innately expressed in terms of pixel display coordinates. These values were converted to degrees of the visual angle (dva) according to the geometry of the recording setup. Although iris images were also captured as part of this collection before the initiation of eye movement recordings, they are not distributed as part of GazeBase. Prior collections including both gaze traces and matching iris images may be found at the following link - 
\href{http://userweb.cs.txstate.edu/~ok11/etpad_v2.html}{http://userweb.cs.txstate.edu/\textasciitilde ok11/etpad\_v2.html}

\subsection*{Calibration and Validation}
A calibration and validation procedure were performed before the recording of each task to ensure data quality. To initiate the calibration process, pupil and corneal reflection thresholds were established. While manual tuning was exclusively used for some initial recordings, the automatic thresholding function of the instrumentation software was ultimately employed to develop initial estimates, with manual fine-tuning performed as required. 

Once threshold parameter values were tuned to ensure successful tracking of the pupil and corneal reflection, a nine-point rectangular grid calibration was performed. During this process, subjects were instructed to fixate at the center of bulls-eye calibration target positioned on a black background. The bulls-eye target consisted of a larger white circle with an approximate diameter of one dva enclosing a small black dot as shown in \figurerefname~\ref{fig:SummaryFig}. The stability of target fixations was monitored by the proctor on the host monitor using a vendor-provided software interface. If necessary, the proctor provided additional instructions to improve image capture quality (i.e.: increase eye opening, etc.).  Once the software determined that the subject had successfully fixated on a target, the calibration process advanced to the next target. Calibration was terminated when a stable fixation had been captured for all nine points in the grid, thereby producing the aforementioned mapping for estimating gaze location. 

A nine-point validation process was subsequently performed to ensure calibration accuracy. Validation points outside of the primary position were disjoint from those utilized in the calibration grid. Validation for each target was manually terminated by the proctor (contrasting from the calibration procedure, which used automatic termination) upon the determination of a stable target fixation. The spatial accuracy of each fixation on the corresponding validation target, hereby referred to as the validation error, was computed after completion of the validation process. Validation error was determined by computing the Euclidean distance between each target and the estimated gaze location. A maximum and average validation error of less than 1.5 and 1.0 dva., respectively, was established as a guideline accuracy criteria for accepting the calibration. However, acceptance of the calibration was ultimately determined by the proctor based upon visual inspection of the discrepancy between the estimated and true target location for each validation point, with additional discretion applied to calibrations failing to meet this quantitative accuracy goal. The calibration protocol was repeated before the recording of each task. 

\subsection*{Task Battery Overview}
A battery of seven tasks were performed during each session of the recording. Tasks were performed in the numbered order described in the following subsections. Acronyms utilized to describe each task within the distributed dataset are defined within each subsection title. 
\subsubsection*{Task 1: Horizontal Saccade Task (HSS) }
The HSS task was designed to elicit visually-guided horizontal saccades of constant amplitude through the periodic displacement of a peripheral target. Subjects were instructed to fixate on the center of the bull's-eye target utilized during calibration. The target was displayed on a black background and was initially placed at the primary gaze position. The target was regularly displaced between two positions located $\pm 15$ dva horizontally from the center of the screen, thereby ideally eliciting a 30 degree horizontal saccade upon each jump displacement. The target's position was maintained for one second between displacements, with 100 transitions occurring during each recording. The proctor notified the subjects of the approximate time remaining within the 100-second recording session at 20 second intervals. An identical stimulus was used for the HSS task across subjects, sessions, and rounds. 



\subsubsection*{Task 2: Video Viewing Task 1 (VD1)}
The VD1 task was designed to elicit natural eye movements occurring during the free-viewing of a cinematic video. Subjects were instructed to watch the first 60 seconds of a trailer for the movie ``The Hobbit: The Desolation of Smaug''. No audio was played during the video clip. The same video segment was used for the VD1 task across subjects and rounds. Due to variability in instrumentation settings, the video stimulus was only displayed for the initial 57 seconds during the second session of each recording. 


\subsubsection*{Task 3: Fixation Task (FXS)}
The FXS task was designed to elicit fixational eye movements through the static presentation of a central fixation target located at the primary gaze position. Subjects were instructed to fixate on the previously described bull's-eye target which was maintained at the center of the display for 15 seconds. The proctor asked that subjects avoid blinking if possible during the duration of the task before initiating the recording. An identical stimulus was used for the FXS task across subjects, sessions, and rounds. 




\subsubsection*{Task 4: Random Saccade Task (RAN)}
The RAN task was designed to elicit visually-guided oblique saccades of variable amplitude through the periodic displacement of a peripheral target. Similar to the HSS task, subjects were instructed to follow the bull's-eye target by fixating at its center. The target was displaced at random locations across the display monitor, ranging from $\pm 15$ and $\pm 9$ dva in the horizontal and vertical directions, respectively. The minimum amplitude between adjacent target displacements was two degrees of the visual angle. Similar to the HSS task, the target was displayed on a black background, with each position maintained for one second. As the trajectory of the target was randomized for each recording iteration, the stimulus varied across subjects, sessions, and rounds. The distribution of target locations was chosen to ensure uniform coverage across the display. 



\subsubsection*{Task 5: Reading Task (TEX)}
The TEX task was designed to capture subjects' eye movements during reading. Subjects were instructed to silently read a passage from the poem ``The Hunting of the Snark'' by Lewis Carroll. The task was automatically terminated after 60 seconds irrespective of the subjects' reading progress. Subjects did not receive explicit instructions of what to do if they finished reading before the end of the 60 second period. Instead, several possible actions were suggested, including rereading the passage.  Because of this ambiguity in instructions, the gaze position towards the end of the recording may vary from the expected per-line pattern typically encountered during reading. Another passage of the poem was displayed as the stimulus for the second session within a given round, with the same pair of sections utilized for all subjects and rounds.

\subsubsection*{Task 6: Balura Game (BLG)}
The BLG task was designed to capture subjects' eye movements while interacting with a gaze-driven gaming environment. During the game, blue and red balls moving at a slow fixed speed were presented on a black background. Subjects were instructed to attempt to remove all red balls from the display area as quickly as possible. Red balls were eliminated when the subject fixated on them, while blue balls could not be eliminated.  Visual feedback was provided to subjects by placing a highlighted border around each ball upon the detected onset of a fixation on the ball. The game was terminated when no additional red balls were remaining. Further details regarding the game may be found at the following link - \href{https://digital.library.txstate.edu/handle/10877/4158}{https://digital.library.txstate.edu/handle/10877/4158}.  

In some instances, steady fixations on red balls did not produce the desired elimination behavior. Based upon this limitation, the proctor instructed subjects to not maintain elongated fixations if the red balls were not eliminating as intended. Instead, subjects were instructed to move their gaze away from the ball, and subsequently re-fixate on the non-eliminating ball. As the initial position and trajectory of each ball was set randomly for each recording, the stimulus varied across subjects, sessions, and rounds. 



\subsubsection*{Task 7: Video Viewing Task 2 (VD2)}
Subjects were instructed to watch the subsequent 60 seconds of the trailer used in the VD1 task. Similar to the VD1 task, no audio was presented to the subjects. The same video was used for the VD2 task across subjects and rounds. Similar to the VD1 task, the duration of the VD2 stimulus in Session 2 was truncated to 57 seconds due to variability in instrumentation settings. 

\section*{Data Records}

GazeBase is available for download on figshare~\cite{griffith_gazebase_2020}. GazeBase is distributed under a \href{https://creativecommons.org/licenses/by/4.0/}{Creative Commons Attribution 4.0 International (CC BY 4.0) license}. Gazebase may be used without restriction for non-commercial applications, with all resulting publications providing citation to this manuscript. All data have been de-identified in accordance with the informed consent provided by subjects. 

GazeBase is organized in a hierarchical directory structure by round, subject, session, and task, respectively. Data records are compressed at the subject folder level. Each task folder contains a single csv file with the following naming convention - `S\_rxxx\_Sy\_tsk', with the relevant parameters for each field summarized in Table \ref{tab:FileConvention}. The first line of each file contains the variable identifiers for each column, which are summarized in Table \ref{tab:FileVariables}. 

\begin{table}[ht]
\centering
\begin{tabular}{lll}
\toprule
Naming Parameter & Definition & Valid Values \\
\midrule
r & Round Number & 1 - 9 \\
xxx & Subject ID & 1 - 335 (excluding incomplete subjects) \\
y & Session Number & 1 - 2 \\
tsk & Task Description & {`HSS', `VD1', `FXS', `RAN', `TEX', `BLG', `VD2'} \\
\bottomrule
\end{tabular}
\caption{Description of file naming convention}
\label{tab:FileConvention}
\end{table}

\begin{table}[ht]
\centering
\begin{tabular}{lll}
\toprule
Variable Identifier & Definition \\
\midrule
n & Timestamp (ms) \\
x & Horizontal Gaze Position (dva) \\
y & Vertical Gaze Position (dva) \\
val & Sample Validity (0 implies valid sample) \\
xT & Horizontal Target Position (dva) (where applicable) \\
yT & Vertical Target Position (dva) (where applicable) \\
\bottomrule
\end{tabular}
\caption{Description of file variables}
\label{tab:FileVariables}
\end{table}

Missing samples within each file are denoted by a non-zero value in the val field, with the corresponding gaze position specified as NaNs. Missing samples result from the failure to extract either the pupil or corneal reflection from the captured image, which may occur under scenarios of blinks or partial occlusions of the eye.  For tasks not employing a target (i.e.: VD1, TEX, BLG, VD2), target entries (i.e.: columns xT and yT) are populated with NaNs at each sample. 


\section*{Technical Validation} \label{sec:Val}
Substantial efforts were undertaken to maintain data quality throughout the experimental design and collection process, initiating with the selection of the recording instrument. The EyeLink 1000 was selected for data collection due to its high spatial accuracy and precision characteristics, which have resulted in its widespread adoption across the research community \cite{nystrom2020tobii}. The EyeLink 1000 is routinely employed as a quality benchmark in the research literature when evaluating emerging camera-based eye tracking sensors (i.e.: \cite{ehinger2019new, lohr2019evaluating,raynowska2018validity}). To ensure adherence to best practices throughout the data collection, all experimental proctors were trained by personnel with considerable prior experience using the device. This expertise was developed during the lab's prior data collections using the EyeLink 1000 (e.g. \href{https://userweb.cs.txstate.edu/~ok11/software.html}{https://userweb.cs.txstate.edu/\textasciitilde ok11/software.html}). 

\begin{figure}[ht]
\centering
\includegraphics[width=0.7\linewidth]{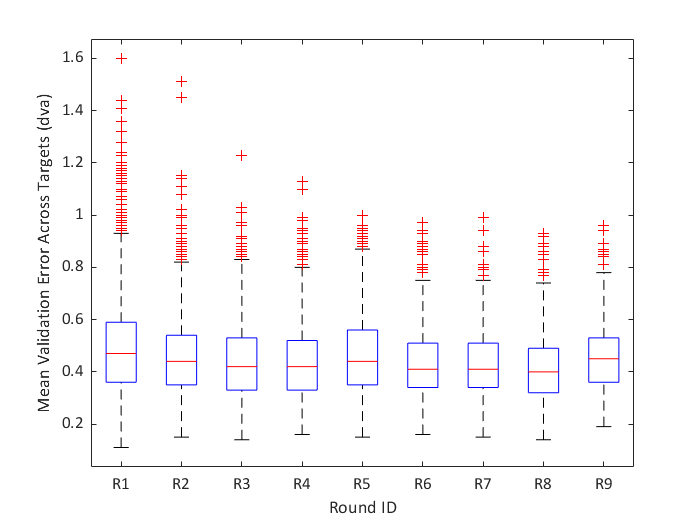}
\caption{Distribution of mean validation error across recordings versus round. The central mark in each box corresponds to the median value, with the lower and upper edges of the box corresponding to the 25th and 75th percentiles of the distribution, respectively. The whiskers extend to the outlier boundaries for each round, which are set at 1.5 times the interquartile range of the distribution above and below the box boundaries. Outliers are marked using the + symbol.}
\label{fig:MeanVal}
\end{figure}

\begin{figure}[ht]
\centering
\includegraphics[width=0.7\linewidth]{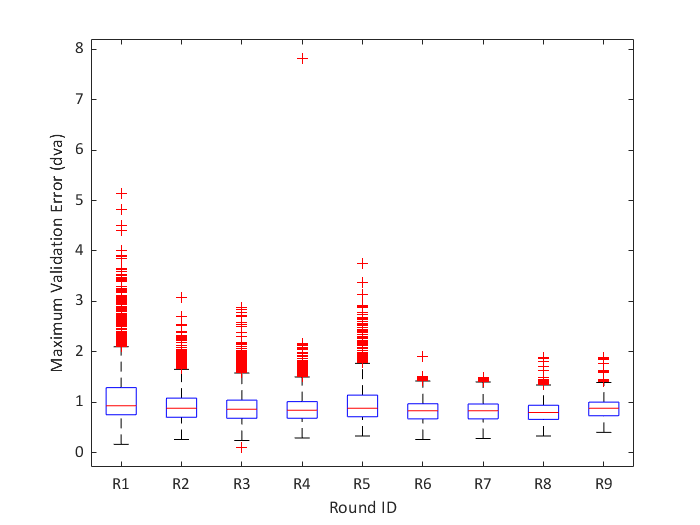}
\caption{Distribution of the maximum validation error across recordings versus round. See Fig.~\ref{fig:MeanVal} for an explanation of box plot parameters.}
\label{fig:MaxVal}
\end{figure}

The experimental protocol was also designed to maximize the quality of the captured data. Namely, subjects were instructed to maintain a stable head position and sufficient opening of the eyelids due to the known relationship between these factors and associated raw eye positional data quality \cite{nystrom2013influence}. A dedicated calibration and validation process was also employed for each task to avoid calibration decay across the collection. Box plots of the distributions of mean and maximum validation errors for each round of recording are presented in \figuresrefname~\ref{fig:MeanVal} and \ref{fig:MaxVal}, respectively. As shown, the median values of the mean validation errors in each round are less than the upper bound of the specified typical spatial accuracy (0.5 dva) of the instrument. Moreover, the significant dispersion of the two metrics, indicating considerable variability in calibration quality across individuals, is consistent with prior observations in the literature\cite{hornof2002cleaning}. 

\section*{Code availability}

The distributed csv files were generated by first converting the edf output files produced by the Eyelink 1000 to a text-based asc file format. These files were subsequently converted to csv files of the specified format using a customized MATLAB script. Data may be extracted from the repository into the target computing environment using traditional csv import functions. 

\bibliography{sample}

\section*{Acknowledgements} 
This work was supported by the National Science Foundation (Award CNS-1250718). Any opinions, findings, conclusions or recommendations expressed in this material are those of the authors and do not necessarily reflect the views of the National Science Foundation. We would also like to thank Alex Karpov and Ionannis Rigas for help in designing the experiments, along with the numerous experimental proctors that contributed throughout the data collection.

\section*{Author contributions statement}
H.G. wrote the manuscript and assisted in data quality assurance,  D.L. provided details regarding experimental protocol based upon his experience as a proctor during later rounds of recording and led data quality assurance efforts, E.A. served as the primary source of information regarding experimental protocol based upon his experience as lead proctor during multiple rounds of recording, O.K. designed the experiments, secured funding, and served as the project manager. All authors reviewed the manuscript.

\end{document}